# Information and Communications Technologies for Sustainable Development Goals: State-of-the-Art, Needs and Perspectives
Jinsong Wu, *Senior Member, IEEE*, Song Guo, *Senior Member, IEEE*, Huawei Huang, William Liu, and Yong Xiang, *Senior Member, IEEE*

*Abstract*—In September 2015, the United Nations General Assembly accepted the 2030 Development Agenda, which has included 92 paragraphs, and the Paragraph 91 defined 17 sustainable development goals (SDGs) and 169 associated targets. The goal of this paper is to discover the correlations among SDGs and information and communications technologies (ICTs). This paper discusses the roles and opportunities that ICTs play in pursuing the SDGs. We identify a number of research gaps to those three pillars, social, economic, and environmental perspectives, of sustainable development. After extensive literature reviews on the SDG-related research initiatives and activities, we find that the majority of contributions to SDGs recognized by the IEEE and ACM research communities have mainly focused on the technical aspects while there are lack of the holistic social good perspectives. Therefore, there are essential and urgent needs to raise the awareness and call for attentions on how to innovate and energize ICTs in order to best assist all nations to achieve the SDGs by 2030.

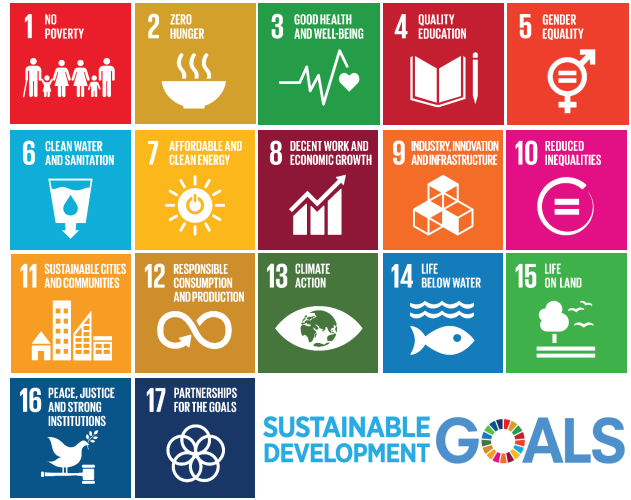

Fig. 1: 17 Sustainable Development Goals adopted by the 2030 agenda for sustainable development [1].

## I. INTRODUCTION

In September 2015, the agenda of sustainable development by 2030 [1] was approved in the United Nations (UN) Sustainable Development Summit held in New York. This summit proposed a new indicator framework, associated with the global and universal indicators, for international cooperations to achieve sustainable development between 2015 and 2030, which, totally, include 17 new Sustainable Development Goals (SDGs), which are shown in Fig. 1, and 169 targets. This new agenda actually continues the well known Millennium Development Goals (MDGs) adopted in 2000 [2], which had been formulated through a Member State-led procedure with wide participation from the major groups and civil public society. The new SDGs framework benefits from the extensive experiences obtained from the implementation of MDGs, and also carries forward partial MDGs unaccomplished and sustains the momentum yielded via addressing the new emerging challenges on equity and urbanization. This further advances the global partnership as well as reflects the continuity and consolidation of MDGs while enabling them to be more sustainable via strengthening the environmental SDGs [3]. Although none of the 17 SDGs particularly refers to Information and Communications Technologies (ICTs), and only several targets mention ICTs and relevant technologies, the 2030 Agenda for Sustainable Development still claims that the ICTs can substantially accelerate the development progress of human beings, and may greatly bridge the digital gaps, so as to construct knowledge communities [1].

The Union of International Telecommunication (ITU) is one of the associated organizations of UN and its mission is to orchestrate the telecommunication services all over the world. ITU has devoted enormous efforts to exhibit the critical roles of ICTs during the establishment of SDGs. ITU has actively participated in the identification of the metrics that are used to measure the SDGs. The derived 17 SDGs and 169 targets are going to stimulate substantial actions over many disciplines in the next fifteen years which are significant for both humanity and our planet. Note that, the SDGs fundamentally target at "5P" including People, Prosperity, Partnership, Peace and Planet as shown in Fig. 2.

ICTs could be the key catalysts to all of the above sustainable development goals, and absolutely crucial to pursue these SDGs. The current research and development of ICT focuses more on exploring the technological challenges, such as the

Jinsong Wu is with the Department of Electrical Engineering, Universidad de Chile, Chile, Email address: wujs@ieee.org.

Song Guo and Huawei Huang are with the Department of Computing, Hong Kong Polytechnic University, Hong Kong, Email addresses: song.guo@polyu.edu.hk, cshwhuang@comp.polyu.edu.hk.

William Liu is with Department of Information Technology and Software Engineering, Auckland University of Technology, New Zealand, Email address: william.liu@aut.ac.nz.

Yong Xiang is with School of Information Technology, Deakin University, Australia, Email address: yxiang@deakin.edu.au.

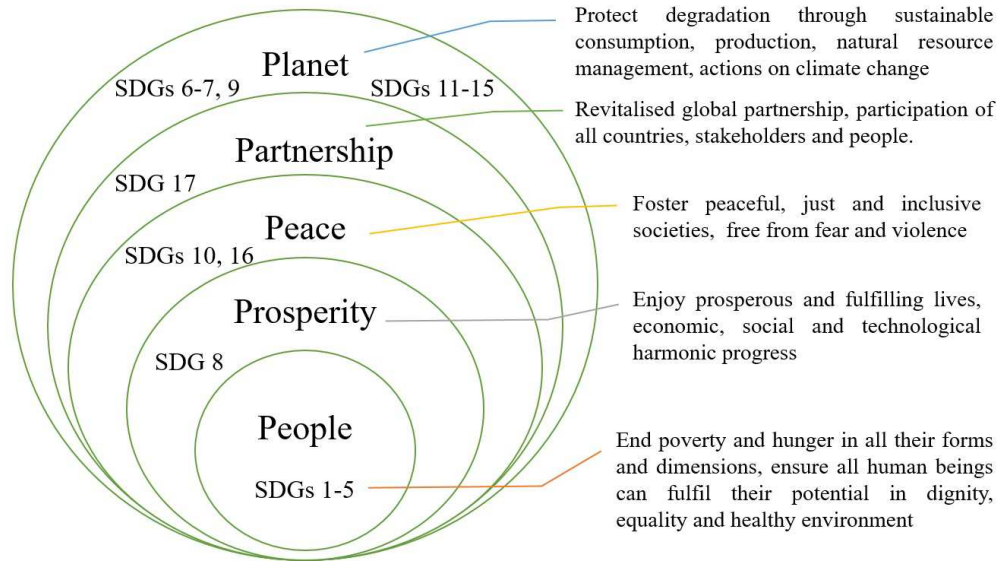

Fig. 2: Five fundamentals of SDGs.

storage capacity, computing speed, computing capabilities, communications, and networking. The research and development toward SDGs heavily requires collaborations among researchers in different disciplines and also extensive communications and collaborations with industry, governments and organizations. Those collaborations and communications may stimulate establishing teamness environments, policies, and regulations as well as building new research cultures of mutual-trust, -recognition and -obligation to each other with high efficiencies.

These post-2015 SDGs have started attracting attentions from ICT domains due to their significant benefits, such as the strong motivations for technological innovation and development, calling for new research topics and also booming research funding opportunities which are all needed to support all nations to tackle those 17 top global issues. This strongly motivates us to investigate the state-of-the-art on the relevance of SDGs to ICT so as to determine how these 17 SDGs have been addressed and what have not been well studied. Although several relevant SDGs to ICT, such as Good Health (SDG 3), Sustainable Cities (SDG 11), and Climate Action (SDG 13), have been well recognized and targeted with extensive literatures reported in the IEEE and ACM digital libraries, some other goals such as SDG 5, SDG 10, as well as the SDG 16 have not been notably aware and addressed by the technical research communities. Our extensive analyses in this paper are to reveal the gaps for those undeserved SDGs so as to raise the attentions of the research communities and also call for their efforts on these SDGs. Thus, the main objectives of this paper are first to summarize the up-to-date research literatures related to SDGs and then to identify the gaps.

The remaining parts of this paper are organized as follows. Section II presents the brief background of SDGs and reviews existing SDGs related ICT research initiatives and activities. Section III, IV, and V discuss the existing literatures and conduct some analyses to identify the gaps in the current works according to three major dimensions of SDGs, respectively. Conclusions are drawn in Section VI, together with laying out the future works.

## II. BACKGROUND

This Section firstly introduces the background of sustainable development and SDGs and then presents the pioneering initiatives and research activities between 2015 and 2016 that are related to SDGs. Finally, we categorize the SDGs based on extensive literature reviews.

### A. The Historical Development of SDGs

Many versions of the definitions for sustainable development have appeared so far. The Brundtland report proposed in 1987 [15] presented the most recognized one, which is described as: "Sustainable development is able to meet the requirements of the current generation and does not have to consume the capability of the future generations. To achieve the sustainable development, people have to find essential ways to turn it from general concept into reality." There have been a number of attempts over the last few decades. For example, the Earth Summit of Rio held in 1992 yielded a common statement known as Agenda 21 [16], which drew a list of actions to be carried out urgently. Later, the World Summit on Sustainable Development held in 2002 clarified the three crucial dimensions of sustainable development, i.e., economical, social and environmental development. These three pillars are independent to each other and mutually supported at the same time. Also, it can be found that the human-respect sustainable development was not much referred to in this summit. Note that, in the early stage, the social perspective of sustainable development had certain ambiguity, which resulted in further discussions and more versions of its definition. Particularly, the original one has aroused more concerns and has been extended to multiple branch directions. In 2011, the human involved sustainable development received special attention in the Human Development Report, which defined the human development as the expansion of its freedom, during which people are proactively capable of determining their life. We can observe from this definition that freedom and capability are characterized notions compared with the basic human requirements. Combining the global environmental issues with MDGs, Griggs et al. [17] proposed six provisional SDGs for people and planet.

TABLE I: Initiatives and activities for SDGs

| Name of Initiative or Activity | Source Information | Vision and Objectives |
| --- | --- | --- |
| SDGs | [4] | It is an online platform for the UN's Division for Sustainable Development (DSD) serving the stakeholders, major groups, and the public, by offering the access to knowledge and information towards sustainable development; |
| SDSN | [5] | It animates the international scientific and technological organisations to solve problems such as design and implementation of SDGs, for sustainable development |
| WSIS - SDG Matrix | [6] | It mainly strengthens the influence of ICTs for sustainable development by connecting the WSIS Action Lines to SDGs. |
| WLL | [7] | An initiative to teach children in over 100 countries about the SDGs. |
| ICSD 2016 | [8] | The target is to bring stakeholders from all layers of society such as united nations, government, academia, NGOs and grassroots organizers together to share evidence-based and practical solutions for sustainable societies. |
| ITU-K Conference 2016 | [9] | This conference recruits academic articles which focus on the developments of SDGs achieved by innovative ICT applications, policy and regulatory. And it particularly studies how international ICT standards benefit the pursuing of SDGs. |
| HCI4S workshop | [10] | This workshop aims to bring people who are working to achieve a sustainable future by contributing to SDGs. |
| DICC4SD workshop | [11] | To complement the related topics of DICC in the main DataCom16 conference, this workshop dedicates itself to enlighten the important role of DICC towards a more sustainable world, by creating research and development, and also innovation for sustainability. |
| MDPI SI on SDGs | [12] | This special issue is to investigate the crucial power in the transformation of ICT to pursuing the SDGs by research communities. |
| Elsevier SI on SDGs | [13] | This special issue is to tackle the challenges that are included under SDG 3: "*Ensure healthy lives and promote well-being for all at all ages*" [1]. |
| ACM SIGCAS | [14] | SIGCAS aims to address the social and ethical consequences of widespread computer usage. The main goals of SIGCAS are to raise awareness about the impact that technology has on society, and to support the efforts of those who are involved in this important work. |

Among different disciplines, the three dimensions of sustainable development and their significances have aroused many different interpretations under various backgrounds. People became more interested in what specifically could be achieved. Institutions working towards sustainable development normally have paid close attention to the related issues. For example, the Millennium Declaration, a UN conference organized in 2000, formulated and reached about 60 SDGs, and the eight famous millennium development goals, which had been adopted globally. However, the wide variety of interpretations of sustainable development in different regions incurred that the definitions, standards and evaluation manners are far away from complying with an identical format, which has degraded the development speed significantly.

Another UN conference on sustainable development, which has been known as Rio+20 (held in Brazil, June 2012) exposed this problem throughout its long preparatory discussions, involving hundreds of stakeholders, when a very large number of issues emerged. These issues are articulated in new ways rather than in the traditional paradigm of the three aforementioned "pillars" of sustainable development. At the outset of the conference, the focus was placed on "green economy", the buzz word of the conference, as well as themes such as food, water, energy, cities, jobs, oceans, and disasters which constituted the main elements of a slim "zero-draft" outcome document, framed within the notions of inclusive human development, equity and human rights. By the time Rio+20 was over, however, this framework was changed again and the consensus achieved in the conference ("The Future We Want") seems to have a much broader scope. Rio+20 also launched a progress which would lead to the review in 2015 on the MDGs adopted in 2000. This agenda has been known as the *Post-2015 development agenda*. Taking advantage of this opportunity, the participants in the conference agreed a new set of objectives, called "Sustainable Development Goals" (SDGs), which deserve the concentrated priority for the conscientious achievement of sustainable development. The agenda document also considered that the future SDGs should be elaborated to be practical and universal to all countries.

Then there was a follow-up report called *Realizing the Future We Want for All*, which was prepared by an appointed Task Team for the previous Secretary-General of the UN, Ban Ki Moon. This report reiterated that the conventional business model needs to be transformed, since the more integrated solutions are in urgent needs to tackle the arising interdependent challenges. This report also set up a vision for the development agenda after 2015, which reflects the most updated thinking as regard to the conceptual framework for sustainable development and serves as a basis to inform the discussion over the next three years. The vision contained in the report is based on three essential values, i.e., human rights, equality and sustainability, and four core dimensions shown as

1) comprehensive social development,
2) comprehensive economic development,
3) concrete environmental sustainability,
4) exact peace and security.

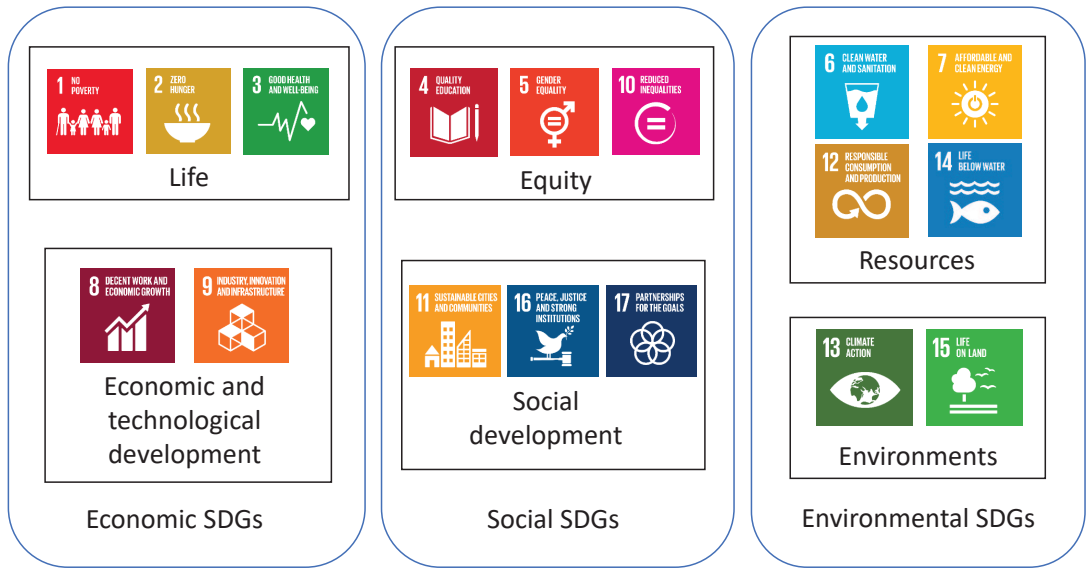

Fig. 3: 17 SDGs are classified into three dimensions according to human needs.

This report also highlighted a set of development enablers and acknowledged that the development standard did not exist. Thus, the proposed agenda advocated that different nations should design policies adapted to local settings under the guidance of the overall vision. The SDGs have been mostly defined based on this principle.

*B. Pioneering Actions on SDG-related Research*

Recently, several pioneering actions and rethinks have appeared for better implementations of SDGs. We identify several representative works as follows. Concentrating on the fundamental moral imperatives of SDGs, Holden et al. [18] proposed a model for SDGs, which could reflect the satisfaction of human needs, the enhancement of social equity and the respect towards environmental constraints. The world health organization (WHO) published the statistics of world health in 2016 [19], which mainly investigated the subjects of health status indicators, world health, health priorities, and other health-related targets. Costanza et al. [20] studied the alternative methods to measure the sustainable wellbeing to stimulate global societal changes, and in particular, proposed a Sustainable Wellbeing Index (SWI) that complements the SDGs. Janowski et al. [21] organized an editorial that explores the correlations and the gap between the SDG implementation and digital government. This editorial study could arise more research attentions for the promotion of SDGs.

In addition, Table I reviews the SDGs related research initiatives and activities after September 2015. It can be observed that there have been some piloting efforts on the SDG topic, but still lacking much attention from the ICT related technical research communities.

*C. The Significance and Categories of SDGs*

The Soviet economist Nikolai Kondratiev, also known as Kondratieff or Kondratyev, was the first to observe the long economic cycle in the modern world economy, who considered the history (from later 1800s to early 1900s) of industrial and technological innovations as three waves, known as Kondratieff Waves, where the start time and end time of each wave are approximations, indicated as some periods [22]. The first wave (1780-1830) was stimulated by stream engine and textile industry; the second one (1830-1880) was represented by railway and steel industries; and electricity and chemistry (1880-1930) dominated the third wave [23]. After the efforts of Nikolai Kondratiev, other researchers continued identifying additionally the following long-waves in the post-World War 1 period as Post-Kondratieff Waves [22]. The development of automobiles and petrochemicals (1930-1970) led to the fourth wave; and recently the ICTs (1970-2010) have led to the fifth great technological age ever since [23]. Generally speaking, Kondratieff Waves and Post-Kondratieff Waves are all called Kondratieff Waves. It is no doubt that each wave mentioned above has accelerated the development and growth of global economy for at least 30 years. There are several recent studies on what are the next wave or the sixth wave. The work [23] thought that the sixth wave would be represented by new materials, particularly those relevant to nanotechnology and biotechnology, (and/or properties of materials) and new processes would play important roles to achieve more friendly environments with fewer resources and less energy, in addition, healthcare technologies would be very important in the the sixth wave. Wilenius and Kurki believed that intelligent technologies would be the key signs of the sixth wave during the period from 2010 to 2050 [24]. The paper [25] even provided a more general prediction that the sixth wave would be the sustainability wave. We think that the predictions of the sixth wave in [23], [24], and [25] could be partially correct.

- The works [23] and [24] have described the features of the sixth wave in the viewpoints of different important technologies in foreseeable future.

- We do agree that the current green and sustainability issues have become much more urgent to the world than those in previous periods of human history. However, we feel that it is too general to use sustainability as the feature of the sixth wave in [25], since we consider sustainability as the universal feature for the whole human history, including the first 5 waves mentioned above, although, in different periods of human history, the levels and manifestations of sustainability could be different, even if people in those period have not had the concepts of sustainability in minds.

According to various perspectives of needs for humanity, the 17 SDGs can be roughly classified into 3 major dimensions [1], social, economic, and environmental sustainable development. Within each dimension, we also category several SDGs related to each other into several smaller groups according to different perspectives of humanity needs [26], such as the self-fulfillment needs, the psychological needs, and the basic needs. The classification is shown in Fig. 3. For each dimension of SDGs, we first discuss the motivation and challenges, and then review the existing related research literatures. Afterwards, the open issues will be summarized.

## III. Economic Sustainability

In this section, we focus on the topics of poverty reduction, food supply, healthcare, economic growth, sustainable industrialization and innovation fostering under the dimension of *Economic*.

### A. Sustainable Life for Humanity

*1) End of Poverty in Life and Technology (SDG 1):* Reducing poverty is still one of the continuing challenges in a long-run during the development of human society in the twenty-first century. The first question is how to determine poverty appropriately. The traditional approaches to identify poverty of an individual and the poor usually based on a rough income-indicator and a poverty line, respectively. However, this simple measurement may lead to an obvious unrealistic classification between the poor and the non-poor over the total population. Thus, efficient approaches that could measure poverty are indeed needed. Firstly, we review the state-of-the-art works on how to measure and reduce the poverty. Othman et al. [27] introduced a composite of multidimensional poverty indicators, named fuzzy index poverty (FIP), which measures poverty by considering the living condition, the ability to possess durable goods and the equivalent income. Yu et al. [28] developed a novel poverty evaluation approach, which particularly combines 10 social-economic factors into an integrated index of poverty. The accuracy of the proposed method has been evaluated in the country level in China, using the nighttime light satellite-data collected from the radiometer suite that captures visible infrared images. On the other hand, there are a growing number of research works on investigating how digital involvements overcome poverty and injustice for communities. For example, Roubaie et al. [29] examined how ICTs alleviate poverty in Muslim countries. To stimulate society development and promote people to participate in economy. This work [29] highlighted the significance of establishing ICT infrastructures based on the fact that empowering nations with ICTs will help to produce and spread new knowledge, thus benefit nation productivity and accelerate the holistic human development. As described in Fig. 4, Figueiredo et al. [30] revealed that the digital inclusion should be realized in two different steps: the first step is to build digital literacy via popularizing the symbiotic computers, such as tablets and smartphones, while the second step is to achieve the professional capability via using the traditional personal computers. It is believed that three administrative moves can mitigate poverty: increase the opportunities of the poor, enhance them with empowerment and guarantee their social security [31]. The reason is that

1) offering opportunity allows the poor people to have accesses to markets, thus providing them the chances to expand their assets;
2) enhancing empowerment inspires state institutions to supply necessary assistances to poor people, especially to the homeless;
3) improving security ensures the poor people to have the ability to cope with risks, such as financial and health risks.

For example, Cecchini et al. [32] reported the practice of poverty reduction in rural India via developing the ICT projects in the manner of enhancing the opportunities of the poor to access to markets, healthcare facilities, and education institutes.

*2) Sustainable Agriculture (SDG 2):* Agriculture is the essential sector to the people living in the rural area in developing countries. Nowadays, the agriculture faces the identical severe challenges in the world. The reasons can be attributed to the following two aspects:

1) The world population keeps fast growth, and the limited agricultural production has to feed the ever growing world citizens.
2) The world agriculture has been constrained and degraded by natural resources, such as water shortages, soil fertility declining, agricultural land production caused by urbanization, and some other negative effects of climate change.

ICTs in agriculture could potentially increase the efficiency, productivity and sustainability through offering information and knowledge. In recent years, smart farming for agriculture [33]–[40] makes a tremendous contribution for food sustainability. For example, using wireless sensor network in farming from independent power source distribution, monitoring irrigation valves and switches operation, and remote area control will effectively improve the production of excellent quality farm products. Several featured studies with ICTs are demonstrated as follows. Awuor et al. [33] highlighted the contribution of ICT to achieve food security and sustainability of agriculture in developing countries. Then, Harris et al. [34] showed a computer vision, in which a new agricultural machinery has been designed to use less pesticides on spray over crops. This study also conveyed that engineering is essential to the supply chain from production to waste disposal. Further, Ren et al. [35] proposed a generic mathematical

**Step 1-** Digital Literacy      **Step 2-** Capacity-building

Symbiotic computer (SC): smartphones and tablets 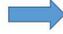 Accomplished with the more traditional personal computer (PC)

Fig. 4: Two steps to achieve digital inclusion for ending poverty in technological aspect.

optimization model, which describes the degradation level of product quality. To control the product quality in food supply chain, the quality degradation of food is particularly considered in the proposed model, in which the wastes incurred by the unqualified items can be reduced. To help understand the concept of new product development (NPD), Driessen et al. [36] developed a framework to empirically study the green NPD. The significance of the proposed integrative framework [36] is that it demonstrates how to coordinate various components of green NPD together and suggests the important role of greenness in chemical and food industries. A notable concept named "seawater greenhouse" [37] has been turned from idea into reality by Charlie Paton, who applied the conventional greenhouse to the seawater. Now the greenhouse under the sea could be used to grow crops. Based on the correlations among water, energy and food security, Hall et al. [38] developed a new model to describe the dynamic global macro-economics. This new model is trying to reveal that the correlation among water, energy and food security is in fact a dynamic system of systems with a certain spatial and temporal complexity. Recently, in order to control farm power distribution and irrigation system, Culibrina et al. [39] proposed a communication methodology of wireless sensor network, which is used to collect environment data and send control commands to turn on or off irrigation system and manipulate power distribution. In particular, using fuzzy logic for decision making in power control distribution for the entire system results in power savings. Singh et al. [40] have developed a solution for agriculture process standards implementation that benefits the quality and quantity of the product. Using a so-called mKRISHI-GAP system, the authors in [40] demonstrated the field experiences in implementation of good agriculture practices certification in Amravati district for orange producer group.

*3) Smart Technologies for Healthy Lives (SDG 3):* Under the era of mobile technologies, the ubiquitous computing and mobile communication technologies have exhibited new concepts of healthcare to industry and governments. Specially, the utilization of information technology and management systems for the healthcare services becomes more and more popular. To build the individual-oriented personalized healthcare service systems, Wang et al. [41] have developed an information service platform based on service computing technologies for public health care. This platform supports almost all conventional personalized services, and even enables the intelligent remote healthcare and guardianship. Solanas et al. [42] studied the smart health using the context-aware applications of mobile health. This is an emerging concept integrated by smart cities and the services provided by mobile healthcare.

To capture the early health changes, Skubic et al. [43] presented a home monitoring system, which embeds sensors in environment and continually observes the behavior patterns of the target. The detected changes of patterns can be used to assess the health changes. Furthermore, Islam et al. [44] reviewed the advances in health care which relies on Internet of Things (IoT) technologies. The up-to-date industrial development, architectures, applications and platforms of IoT based healthcare have been introduced. In particular, the work [44] also included security related taxonomies in the context of healthcare.

*4) Education Promoting and Learning Opportunities Brought by ICT Technologies (SDG 4):* Nowadays, educators could deliver knowledge to students more conveniently using the Internet and media-rich web applications, compared with the conventional classroom based approaches performed in the face-to-face manner. These new technologies enable the students who cannot attend classes receive the same education opportunities. Chou et al. [45] discussed the distant learning education cooperated with the computer networks. Based on using information technology, Latchman et al. [46] presented a model to assist the learning experience for both school students and the asynchronous-learning students. To improve the performance of students, the work [47] studied m-learning techniques and developed 13 smartphone based applications to help refine the functionality of m-learning. Later, Tsai et al. [48] proposed a web 2.0 based platform for sharing course assignments, research and social issues in a university class. Oriti et al. [49] created a doubly fed induction machine (DFIM) for supporting the distance learning education. Goldweber et al. [50] presented a social-good enhanced framework for computer science students. Through providing social good related assignments, and examples of introductory computing projects, students can be conveyed the social relevance and positive impact of potential good outcomes.

In recent years, the Ubiquitous Learning Environments (ULEs) have been proposed for the sake of that the mobile devices are widely adopted in the scope of education. Thus, technological and social ecologies have been integrated into the context of eduction. This trend also has brought additional challenges and burdens to teachers when they participate in such a complex education system equipped with technological opportunities. Thereby, Cristobal et al. [51] developed a system called Group Learning Unified Environment with Pedagogical Scripting and Augmented Reality support (GLUEPS-AR), which is designed to alleviate the burdens of teachers under the ULEs based learning scenarios.

## B. Economic and Technological Development

*1) Sustainable Economic Growth and Productive Employment and Decent Work (SDG 8):* In the twenty-first century, with the environmental and social responsibility being sensed widely, it becomes an inevitable option for industrial enterprises to adopt the concepts of systematic development and implement the economic-social-ecological development. Thus, industrial enterprises not only have to pay attention to the productivity and investment efficiency, but also need to take the sustainable development issues into account. For example, by considering the sustainable development, Manion et al. [52] have studied the philosophy of engineering ethics, aiming to help realize all facets of sustainable technology. To achieve the sustainable development and realize an economic-harmonious system for industrial enterprises, Li et al. [53] have explored the social-ecological-Then, Manish et al. [54] have emphasized on the sustainable collaboration model between academia and industry, the design options and challenges have been also discussed as well.

*2) Sustainable Industrialization and Foster Innovation (SDG 9):* To realize the sustainable development for industry, one big challenge is to utilize the finite land resources in a reasonable and planned manner. To this end, Dudukovic et al. [55] used the spatial-decision-support system techniques based on geographic information systems, to help accomplish the proper siting for newly planned industries. The proposed software tool is great useful to the developing countries, where the sustainable industrialization is still in their premature stages. Sotoudeh et al. [56] presented a whitepaper which describes the significance of technologies on sustainable development. For example, some strategies supporting cleaner production, and the Environmental Technologies Action Plan (ETAP) of European Union. It is also pointed out that to achieve sustainable development requires not only adopting the environment-friendly technologies for economic growth. Jofre et al. [57] studied the dynamic mechanisms that affect the evolution of industrial products when adapting to the changing environments. Then, authors also discussed the significance of evolutionary theory on environmental innovation. Finally, an incremental innovation based strategy called Eco-evolution has been proposed. Furthermore, aiming at understanding the procedures, functionalities and roles in the end-to-end innovation, a framework was presented in [58] as the implementation of the focused concept, whose feasibility has been validated via the application showcase in renewable energy division. Since 1990, the fast developed informatization has made great contributions to industrialization. Using the theories capturing the correlations between sustainable development and economic growth, Wei et al. [59] investigated how to test the convergence performance of informatization and industrialization, which was concluded using some recommendations for promoting full convergence and transforming the conventional industries. Then, Golovatchev et al. [60] summarized their innovation projects experience of several years in the stimulation of innovation. A strategic tool named technology and innovation radar was elaborated in the processes of identification and evaluation for newly startups. Over the last decade, companies have devoted many efforts to sustainable strategies, which have been found that innovations are effective when applying to business model. Aguilar et al. [61] compared two models found in recent literature, the sustainability in the project life cycle model (SPLC) and the sustainable business model (SBM), to find answers on identifying how innovation management benefits economic performance. In addition, people are searching new solutions that yield more sustainable innovations when facing the challenges of various global, economic and ecological demands. Koskela et al. [62] have studied whether people can gain more sustainable innovations by learning from experiences. To achieve sustainable innovation, the purchase innovative products and services is necessary to consumers. Based on this fact, Frank et al. [63] have identified the determinants that impact the sustainable innovativeness of consumers across different countries. To gain the insight of the distributed software development, Jalali et al. [64] have systematically reviewed the literature published in the last decade that combining agility and global software engineering.

## C. Open Issues that Need Embedded Sustainability into the Engineering Design

Although existing studies have improved the sustainability of human's economic life, we summarize the following open issues that still need to be addressed. From the aforementioned works related to poverty reduction, we can see that the education on effective ICT usage is a critical key that should be strengthen to reduce poverty further. Although engineers have made a lot of contributions to the economic development and the quality of human life worldwide, some unintended side effects caused by engineering works brought enormous degradation to both natural and social systems, whose consequences have been criticized in the past decades. Therefore, it is in urgent needs that the engineers should rethink the new way to deliver projects to tackle both technical and social challenges, and take the sustainability into account simultaneously, especially to the engineers who work in the underdeveloped regions. Note that, the considered technique and non-technique aspects may be relevant to the wide range sectors, such as water providing, sanitation guarantee, healthcare, power generation, siting choosing, infrastructure design, food science and communications, and so on [65]. A challenging issue on the worldwide urbanization process is to gather larger population for cities and offer citizens efficient services. Furthermore, the Internet of Things makes modern cities smarter than the conventional ones. One of the significant application domains for IoT is healthcare. The smart IoT that reform the future healthcare with the considerations of technological, economic, energy-efficient and social sustainability will become one of next promising research directions for both academia and industries. On the other hand, organizations worldwide are paying more attention on the sustainability of their business, aiming to increase economic growth [66]. The professional IT purchasers are demanding reassurance that the IT products are devised by considering the social, economic and environmental sustainable features.

Therefore, it is significant that the IT engineers can handle all the factors reflecting the social responsibility, reputations and values of customers when designing their products.

In view of the historical tasks to realize sustainable industrialization, for the development of recycling economy, it is necessary to solve the constraint of limited resource [67]. Thus, led by the sustainable development concept, the pace of technological innovation must be accelerated via developing high and green technologies. It can be seen that the transforming the traditional industries into the ones enabling recycling economy is a significant issue in the way to realize industrialization.

In the contexts of growing globalized competition, it is no doubt that sustainable innovation becomes increasingly important. The fundamental research organizations usually reflect the innovative strength of a nation. They also may undertake the important responsibilities in transforming innovations into market profits and maintaining competitive advantages. Thus, the applications of innovation concepts to sustain their innovation strength are in urgent needs for these organizations.

## IV. Social Sustainability

Nowadays, the disparities over the opportunity, wealth and power in our society become significant. It is believed that the gender inequity plays a pivotal role behind, and the unemployment, typically youth unemployment, is a main reason. On the other hand, the inequities within and among countries are also increasing, leading to billions of world citizens still attached to poverty. Therefore, the elimination of gender inequity and the reduction of inequity among countries are crucial targets on the way to sustainable development in the future human society. Thus, this section mainly discusses the equity issues in the respects of gender and country as well as the relevant open issues to pursue the sustainable social development.

### A. Pursue Gender Equity and Reduce Inequity among Nations

*1) Achieve Gender Equity (SDG 5):* Long et al. [68] studied the gender issues in depth by setting the background of case study in the undergraduate education for the major of computer science in China. This article discussed various existing gender issues, such as the correlations between gender divergence of employment and subject major, the imbalance of gender in the information companies, and so on. This study unveiled that the social conventionality and industry culture determine the source of gender issues for computer science education in China. Historically, women were excluded in the technology-intensive fields. Cohen et al. in [69] presented that the gender gap tends to be reduced in some online activities. In contrast, based on Davis Technology Acceptance Model (DTAM), Jong et al. [70] proposed a new model which captures multiple factors including social-economic class, gender role, gender equality, and self-imaging variables. Based on the observation of female students in different college years, the results have shown that significant differences exist in gender roles. In the face-analysis field, the classifications of both gender and ethnicity are big challenges. Some human features are only useful for one application aspect. For example, skin colors are only suitable for the identifications of ethnicity. Some other features are common to both application aspects, such as face profile. However, the mutual impacts between classifications of ethnicity and gender are still an open issue. Therefore, Farinella et al. [71] studied this subject about whether the classifications of ethnicity and gender affect each other. As a result, three classification algorithms were proposed. The experimental results implied that these two aspects do not dependent to each other during the classification processes. Furthermore, in the field of speaker recognition systems, the best accuracy can be attained when gender dependent features and gender trials are known. However, the labels used to identify gender are normally not available in real-world applications. Cumani et al. [72] have developed a completely gender label-free system based on machine learning mechanism.

Social activities are important for different genders. In recent years social activities are highly supported by communications networks, which are called social networks or social media. The work [73] studied gender identification on social networks via reposting behaviors using the combination of statistics and sociology. Another paper [74] proposed to jointly use the Latent Semantic Indexing (LSI) method and K-nearest neighbor algorithm (KNN) to predict genders through actual data of blog pages.

*2) Reduction of Economic Inequality in Country Layer (SDG 10):* Economic inequality can be estimated by Gini coefficients [75] under most circumstances. However, the Gini coefficients cannot be used to explain the effects of economic inequality on society. To capture the effects of economic inequality, Lu et al. [76] proposed to use membership function in the measurement of both positive and negative effects of relative poverty. Finally the attitude-index based approach was derived. Then, using the proposed theory, authors investigated the influence of economic inequality with the background of multiple typical districts. The economic behaviors existing in economic system, such as the price determination of products, are the consequences of economic equilibrium, which plays the role as the knob to adjust the market prices of both supply and demand sectors. Wang et al. [77] developed a basis theory to retrieve the economic equilibrium in the finite continuous topological spaces for economic departments. In 1955, it was conjectured that the economic progress normally goes with growing inequality in the early phases of industrialization. The inequality fluctuation will shrink following the progress of industrialization and the sufficiently distributed development benefits. However, the observed inequality patterns shown between 1970s and 1980s did not comply with this theory. To find the reasonable explanations behind the phenomena, Conceicao et al. [78] have analyzed the inequality details during the period 1970s to 1990s presented in the economic development countries. The findings have shown that the correlation between the economic growth and the economic inequality does not hold in the countries with highly developed economy.

More recent literature on economic inequality can be recognized as follows. Jiang et al. [79] have studied new models to describe the stability of economy. Based on the approach of linear matrix inequality, a discrete-time system controller with the functionality of state-feedback has been derived. Then, to depict the development inequality with spatial features in Xinjiang district of China, Liu et al. [80] have exploited the spatial visualization technique, in which the geo-statistics was specially viewed as a metric to estimate the inequality level. Recently, Park et al. [81] have concentrated on a problem of political economy. The correlations between the tax reduction induced by certain politic policies and the economic inequality has been studied. The results of this investigation have suggested that if most citizens would have been well-informed with the inequality behind the political policies, the tax-reduction revolution would be voted to be disproved. Those policies proposed by scheming politicians and aiming to enlarge the inequality would become invalid.

*B. Social Development Calls for Sustainability*

*1) Sustainable Cities (SDG 11):* Sustainable cities are certainly in the spotlight when people mention the sustainable future. The reasons can be attributed to the following two-folds: (1) there will be around 70% of the world population living in urban areas by 2030, and (2) cities may be responsible for over 70% of global greenhouse-gas emissions [82].

The recent research efforts focusing on sustainable cities are reviewed as follows. Zen et al. [83] presented a proposal of sustainable energy indicators for cities, based on indicators in five dimensions, i.e., environmental, economic, social, territorial and political. The analysis of these five dimensions could help policy-makers foster sustainable energy development in the future cities. The necessity and usefulness of making cities smarter is absolutely clear to citizens. However, it is not easy to launch the businesses on smart cities, since making the value of this concept has not yet apparently observed. Thus, Vilajosana et al. [84] investigated the reasons behind this phenomenon, and finally made a very meaningful conclusion that it is necessary for smart cities to be built on a basis established via exploiting the ICT based big data technologies. For the resource-exhausted cities, sustainable development must be brought to the table immediately to pursue both economic and social sustainability, for which the urban spatial structure (USS) is an essential tool. Based on the knowledge of USS, Lin et al. [85] studied the sustainability of cities in three respects including society, economy and environment. Concerning the attitude to treat the role of information in the implementation of sustainability, Cosgrave et al. [86] have provided surprising viewpoints that it may be harmful to the urban planning to predict the future due to the disruptive possibility to sustainability. Internet of Things (IoT) have been widely considered as the future fundamental platform to achieve the sustainable smart cities. Vlacheas et al. [87] identified the obstacles in the course of reaching the goal. Some notable issues are summarized. For example, the nonuniformity of the connected things brings barriers to handling the variety of data generated by IoT devices. This paper [87] also unveiled a promising road to achieve the smart cities, i.e., developing the IoT with self-reconfigurable functions.

Krishnaswamy et al. [88] proposed a tool for sustainable cities named TripleRM, which can be utilized to manage risk, resource and resilience of cities for long term sustainability. For example, using this proposed model, the citizens around the world could enable their cities via proactive planning to cope with the natural disasters, social, and economic damages. Many projects towards advocating the renewable energy have been launched by European Commission, specially for the large cities. For example, the project POLYCITY [89] has demonstrated an attention-attracting energy policy, which has been carried out in three European cities. Then, another data-mining based project named SMARTY (SMARt Transport for sustainable citY) [90] has been initiated in the Tuscany region of Italy, aiming to stimulate the innovation devoted to sustainable transportation system serving the future smart cities. Nowadays, the increasing natural disasters, such as extreme hurricanes, floods, heat waves and droughts, are triggering the alarms that the climate changes have led to significant challenges to the physical, economical and environmental life of humanity. It has been believed that earth observation (EO) technologies are able to effectively support alleviating the effects of climate changes [91]. Accordingly, to bridge the gap between the existing up-to-date approaches to cope with climate changes and the urgent requirements, European Union FP7 Program set up the project of DECUMANUS (DEvelopment and Consolidation of geo-spatial sUstainability services for adaptation and environmental and cliMAte chaNge Urban impactS) [91]. Specially, this project has provided options for the managers of city planning with EO-based geo-spatial products, which offer sustainable services to tackle the environmental and climate changes. Finally, the URBANETS (Sustainable Management of Urban Networks with the use of ICT) project [92] has been viewed as a good example of using ICT to build the sustainable management of urban networks for two important Italian cities, Brindisi and Gallipoli.

*2) Sustainable Consumption and Production (SDG 12):* The concept of sustainable consumption and production (SP&C) was proposed in 1992 by the Earth Summit in Rio. Then, this concept has been invoked by the government when making policies for promoting public services. Dillon et al. [93] discussed some policies of organizations and industries that pursue SP&C, and offered suggestions on the manner of realizing the sustainable goals. Ludemann et al. [94] implemented an archetype of sustainable factory for solar cells, which ensures high productivity, quality and assurance. To find the balance among the three aforementioned dimensions of sustainability in Japan, Ueda et al. [95] proposed a tool medium named "Green Point Card System" for sustainable consumption. In view of the growing conflicts between the environment resources and the demands of society, unlike the traditional style of consumption, Xu et al. [96] proposed a conceptual model removing the waste phase and using recycle phase for sustainable consumption. Then, Sotoudeh et al. [97] presented an important official statement of European Union, which described the significance of environmentally friendly technologies in the respect of sustainable development. In today's society, the demands of energy consumption keep growing, and the power generations mainly rely on the fossil fuels, such as gas and petrol, which are unsustainable resources. To monitor the demands of energy, Nambi et al. [98] explored the plug-level meter based appliance for power and resource consumption in house. Wigstrom et al. [99] focused on the scenario of manufacturing industry, and proposed mechanisms of energy saving. Recently, Brundage et al. [100] broke down the structure of energy consumed by production, aiming to discover the real reason behind the low efficiency of energy utilization in modern manufacturing facilities. Liotta et al. [101] proposed a methodological framework, which coordinates the production and the system of transportation. Particularly, the demand uncertainties under the proposed framework could be well dealt with.

*3) Sustainable Development, Provide Access to Justice (SDG 16):* Thanks to the technology advances and the progress brought by the Information & Knowledge Society, the electronic-justice emerges as new tools that provide people more convenient access to justice than the conventional manner. Xia et al. [102] have witnessed the establishment of electronic-government that offers society great opportunities to receive information justice services. Recently, electronic-justice emerges as a necessary sector of the electronic-administration, especially since it contributes a lot to the constitution of modern justice administration system and the openness in public services. Cano et al. [103] have shown an innovative vision of electronic-justice based on the concept of citizen-centric in the development of ICT in legal issues. Lista et al. [104] have presented a practical web platform to supporting management and delivery of information about justice services offered for Colombia citizens. Based on group engagement model, Li et al. [105] built a model that describes the relationship among the justice, work engagement and organizational identification. Liu et al. [106] mentioned that the Chinese Characteristics Environmental Justice Ethics (CCEJE) has been built up both on the basis of John Rawls Theory of justice and the reality of China.

*4) Strengthen the Means of Implementation and Revitalize the Global Partnership for Sustainable Development (SDG 17):* First, Mayende et al. [107] envisioned the advantages brought via adopting the mobile and its ambient products and technologies during the collaboration of facilitating school education. Then, authors argued that this manner of inclusion of mobile technologies benefits the sustainable education in the undeveloped countries. Mann et al. [108] examined the relationship between collaboration and sustainability. The collaboration has been advocated to achieve the sustainability. Later, Ruokolainen et al. [109] studied the sustainability of service ecosystems, and proposed a service ecosystem architecture framework, which can help analyze and even design the sustainability of ecosystem via offering the stakeholders the solutions to eliminating their concerns. Then, Jacovella et al. [110] reviewed the critical issues on collective awareness platforms, and showed how collective awareness help to improve society structures. Vecelak et al. [111] studied how to handle the critical difficulties that complicate the long-term research and sustainability when more teams collaborate in a number of research areas simultaneously. Recently, in order to show the merit of collaboration between organizations and inspire others to do that similarly, Hankel et al. [112] unveiled the experience to sustainability through their work accomplished in a collaborated team.

*C. Open Issues to Enhance the Social Sustainability*

The mentioned articles related to reduction of inequity towards both gender and countries mainly discuss how to evaluate such inequity. However, how to alleviate inequities is still open issues that need much more efforts. As the urban population increases quickly, we need to reconsider an emerging issue that how to create a modern city with less stressful and more creative environments. Live ability and life quality are viewed as key factors while designing the long-term energy, water, pollution and waste managing systems [113]. New approaches are expected to be proposed by taking these goals into considerations. Furthermore, it is widely accepted that the bottom-up designs [114] should be adopted to realize the future ICT-enabled sustainable cities. This arise a challenge for city leaders that systematic thinking needs to be applied when making the core principles of future sustainable cities. On the other hand, deeper collaborations between information technologies and justice systems could spawn various electronic-justice systems, which will empower lawyers and judges when exploring documentary evidence [115]. Therefore, more ICT-based tools and techniques with enhanced knowledge-discovery are expected to be developed in the future.

Since many critical practices in the context of social sustainability are global issues, such as globalized urbanization, implementation of smart cities, sustainable energy production and consumption. In addition, the collective intelligence [110] can be used to pursue the improvement of holistic society based on the collective efforts contributed from individuals. Therefore, many more collective awareness global partnerships are necessary for the sustainable development of future human's society.

V. ENVIRONMENTAL SUSTAINABILITY

In this section, we discuss the state-of-the-art studies that aim to conserve the environmental resources and achieve the environmental sustainability.

*A. Sustainable Natural Resources for Our Planet*

*1) Sustainable Management of Water and Sanitation (SDG 6):* As reported in [116] that a large portion of world population have difficulties in obtaining accesses to the very basic services for human. For example, some people still hardly receive safe drinking water or the sanitation facilities. This potentially increases the political instability and worsens the cycle of poverty. Furthermore, with nearly 1000 children dying per day due to the waterborne diseases [117], drinking water crisis is ranked the first on the global risks by World Economic Forum, 2015. Of those who face the problem 82% are from rural areas. To make changes on conventional water supply, Kedia et al. [118] proposed different sensor-cloud based systems to improve the water quality. For sharing the information about the requirement of water supply and sanitation services, Chisholm et al. [119] developed a web-based system called DevClear that enables organizations to meet the needs in communities. Moreover, Kodjovi et al. [120] theoretically analyzed the influence of user participation during the quality guarantee and regulation delivery of the services of water and sanitation.

*2) Access to Affordable, Reliable, Sustainable and Modern Energy - SDG 7:* To achieve the sustainable development of society, one of the most important factors is the supply of energy resources that need to be fully sustainable [121]. Matsuhashi et al. [122] has clarified the key points in realizing the sustainable energy supply under the restriction of $CO_2$ emissions. Then, Ilic et al. [123] studied the essentials of smart grid technology in the ICT enabled sustainable energy services. A general framework of socio-ecological energy system was proposed. Later, Camacho et al. [124] presented the development of Li-ion batteries based Electronic-Vehicle (EV) technologies. EV technologies are still in the premature stage. A lot of designs and products are in the strong needs via the Original Equipment Manufacturers (OEMs), EV industries and research institutions. Therefore, this new battery product can provide the foundation for EV industries.

### B. Environments Call for Sustainability

*1) Urgent Actions to Combat Climate Changes (SDG 13):* The beginning of this century has witnessed quite a number of impacts of climate changes, such as iceberg melting, sea level rising caused by global warming, extreme drought occurred frequently, and drastic ecosystem change. Many urgent actions need to be devoted to combating the environmental degradation due to climate changes.

We list several existing state-of-the-art efforts as follows. Mills et al. [125] discussed the strategies to mitigating the greenhouse-gas emissions, and other aerosol emissions that do harm to the balance of atmosphere. Camyab et al. [126] have informed and educated the public about what they could do to respond to climate changes, such as boosting the international cooperation to develop low-carbon technology and improve the energy efficiency. Then, Cooper et al. [127] conducted a monitoring program for the northern bio-region of Canada to assess the climate change indicators, such as the temperature of soil and air, the duration of snow cover, the decomposition rates, and glacier scale, etc. Parkpoom et al. [128] studied how the changing climate affects Thailand's demands in daily, seasonal, and long-term periods. Janicke et al. [129] proposed to use a holistic analysis approach to enable users to visually observe the global climate changes, using a 250-year spanning dataset that consists of various variables. To better understand the climate change, Wang et al. [130] have investigated several computational issues based on the modern high-fidelity climate models. The project of Ocean-Color Climate-Change Initiative (OC-CCI) [131] was set up to study the climate change by yielding the uniform ocean-color products with long-term and error-characterized time series. And Lo et al. [132] studied the green hotel and to help people understand the surveyors's cognitions of climate change. Recently, Liestol et al. [133] realized a situated simulation that visualizes the climate change using augmented reality technologies, which can effectively convey the essential information of climate change to the public. Hagenlocher et al. [134] studied the population change of subsaharan Africa, where the extreme climate events occur frequently. You et al. [135] have quantified the effect of climate change and considered other sources of uncertainties at the same time.

Particularly, a notable recent event that illustrates the determination of the world to mitigate global warming refers to the Paris Agreement [136], which is an agreement belonging to a part of the United Nations Framework Convention on Climate Change (UNFCCC) international environmental treaty. The aims of this agreement are to deal with the greenhouse gas emissions and to enhance the implementation of climate-resilient development. There have been 195 countries and nations have signed this agreement, and some of them have already taken actions to perform their promises. For example, the minister of France's environment has announced in July 2017 that France decides to ban the fossil fuels including petrol and diesel by 2040, and a large budget will be ratified to the development of high-efficiency energy.

*2) Sustainable Development of Oceans and Marine Resources (SDG 14):* Ocean system provide a number of critical functionalities such as offering food and energy resources, supporting trade and commerce, maintaining biodiversity within the earth system [137]. It also intensively couples with the weather and climate variations. Thus, it is quite significant to maintain the sustainable development of both oceans and marine resources. Walpert et al. [138] reported the oil spill off event occurred in the cost of Louisiana in 2010. This disaster caused enormous damages to coastal wetlands and taught people lessons again with the high risks on the operations of offshore drilling. To conserve ocean and marine resources, many initiatives have been established in recent years. For example, a new project plan for national marine science named the Canadian Healthy Oceans Network [139], which recruited researchers for making technical handbook and guidance for sustainability while exploiting the marine biodiversity resources in three oceans of Canada. This program is mainly to address the relationship between the habitat diversity and biodiversity patterns, and to develop approaches that bridge science and sustainable policy. With the vision to effectively manage the sustainable development of coastal areas, and protect the ocean life at the Placentia Bay, an initiative named SmartBay [140] has been designed as a user-driven system for ocean-observation. This system was launched by the Fisheries and Marine Institute of Memorial University of Newfoundland, and has served the Placentia Bay, Newfoundland with the information needed by users including fishermen, oil industry, marine transportation, municipalities and the residents who live there. As the first explorer to operate the fish farm in the open ocean off Kona in Hawaii, Kona Blue [141] is leading the innovative mariculture, which, however, is subject to various constraints related to sustainability. For improving coordination in ocean observations, Pearlman et al. [137] reported the early phases of some efforts made during the organizing a multidisciplinary team, named Ocean Research Coordination Network (ORCN), which has been sponsored by National Science Foundation (NSF) and launched for building an efficient platform for the cooperation performed in the community of ocean research.

In addition, some ocean-observing systems can be summarized as follows. Aiming to improve the competitiveness of the aquaculture industry, Cater et al. [142] have developed a sensor based ocean observing systems under the Smart Ocean Sensors Project. Waldmann et al. [143] built a framework of ocean observing infrastructure to improve the transatlantic cooperation. Buckley et al. [144] established an integrated ocean observing and prediction system, which was designed to build a model for predicting the storm, surge and flooding in the estuarine and coastal areas of Carolinas based on the monitored ocean conditions.

On the other hand, the ecosystem-based management has attracted much attention in modern marine environment studies. For example, Kononen et al. [145] summarized the contribution of the marine research conducted to maintain and improve the ecosystem of Baltic Sea. Biswas et al. [146] studied the influence incurred by climate change, specially the water-temperature change, on the fish catches in sustainable management of world fishery. Recently, Ferreira et al. [147] reported the TURTLE project that explores the underwater subsystems. This study could reduce the operational expenditure, increase the capability of human activities, and benefit the sustainable operations in the sea bottom.

*3) Protect, Restore and Promote Sustainable Use of Terrestrial Ecosystems (Goal 15):* The forest ecosystem is the most typical and complicated in biodiversity in land ecosystems. Qin et al. [148] analyzed the entropy change on forest ecosystem under the bearing capacity and coordinated control mechanism. Hoekman et al. [149] studied the sustainable forest management in Indonesia using the advanced radar techniques. Wood et al. [150] introduced the space-based sustainability of forests in Canada based on the technologies of earth observation. Using TerraSAR-X imagery, Molinier et al. [151] investigated the forest cutting monitoring in images over Kuortane, Finland. Recently, Mahabir et al. [152] examined the climate change and sustainable forest management via adopting the geospatial technologies.

Land degradation is a critical issue in the sustainable ecosystem. The related notable research effort can be recognized as the follows. The study [153] demonstrated the usage of remote sensing data and how the data was used to assess the land degradation. Then, Li et al. [154] discussed and analyzed the environmental impact of tourism exploration to the desert areas. Authors strived to find approaches to achieve the sustainability for the desert tourism development. Zhang et al. [155] described the necessity of biodiversity protection in land use, under the concern of protecting the biodiversity in the traditional land-use. Recently, Feng et al. [156] analyzed the land degradation in Zoige wetland ecosystem to find the land degradation pattern via analyzing the MODIS-NDVI (Normalize Difference Vegetation Index) data.

Remote-sensing has been viewed as an useful tool in the natural resource studies. For example, to sustainably develop the land and water resources, Rao et al. [157] developed a remote sensing based approach, which integrated with a geographic information system. Asner et al. [158] studied the desertification of drylands in Argentina applying the imaging spectroscopy. The photosynthetic vegetation (PV), non-PV, and the cover ratio of bare soil were tested and analyzed, based on the measured earth-observing data collected by the airborne imaging spectrometer. Wu et al. [159] presented the approach exploiting signal processing to model the herbaceous biomass using the data sensed by the airborne-based laser scanners.

In addition, based on Geographic Information System (GIS), Li et al. [160] investigated the ecological function regionalization and planning for Shangri-La County, China using graphs superposition. Silva et al. [161] have reported a biodiversity data management framework, which could efficiently integrate and retrieve large volume of biotic and abiotic data. Recently, Martinis et al. [162] studied the ecological protection, the sustainable environmental conservation and the socio-economic development in Natura 2000 areas in Ionian Islands, specifically at Zakynthos and Strofades. Shyam et al. [163] studied the water resource management based on ICT technology. As a result, an Internet Geographic Information System application has been developed.

*C. Open Issues for Environmental Sustainability*

For the long-term efforts devoted to the reduction of $CO_2$ emissions, the marine energy resources, such as wave, tide and offshore wind, have been considered as the reliable source of the generation of renewable energy [164]. Therefore, the use of marine energy is viewed as a promising direction for future sustainable energy supply for human. When people are trying to discover new sustainable energy resources, increasing the utilizing efficiency of both the current and the potential energy resources also should be a major concern.

Europe is in the leading position with respect to abandoning the traditional energy sources (i.e., gas, coal and oil based fossil fuels), and adopting the renewable energies. In contrast, the other big energy-consumption countries, such as United States and China, still move slowly in the way to pursue renewable energy. Furthermore, European Union has set ambitious goals for 2020 to improve the production of renewable energy and energy efficiency, while remaining efforts to reduce the greenhouse gas emission [165]. Besides, the evidence of climate changes shows that the human activities are the main reasons, such as the massive use of fossil fuels. This further stimulates the international community, especially the large nations, to consider setting the new energy models into their national development agendas. On the other hand, the widely adoption of smart devices in homes appears as a promising approach to realize the intelligent homes under the concept of Internet of Things. Because with the integration with smart devices, the residential electrical power systems can work in a more cleaner and sustainable energy mode.

Climate change is viewed as one of the featured global issues in 21st century [166]. To adequately address climate change, people have to shift from the statistical analysis to scientific insights to understand the climate science. In consequence, the novel data-science approaches that can handle the spatio-temporal and physical nature of climate phenomena

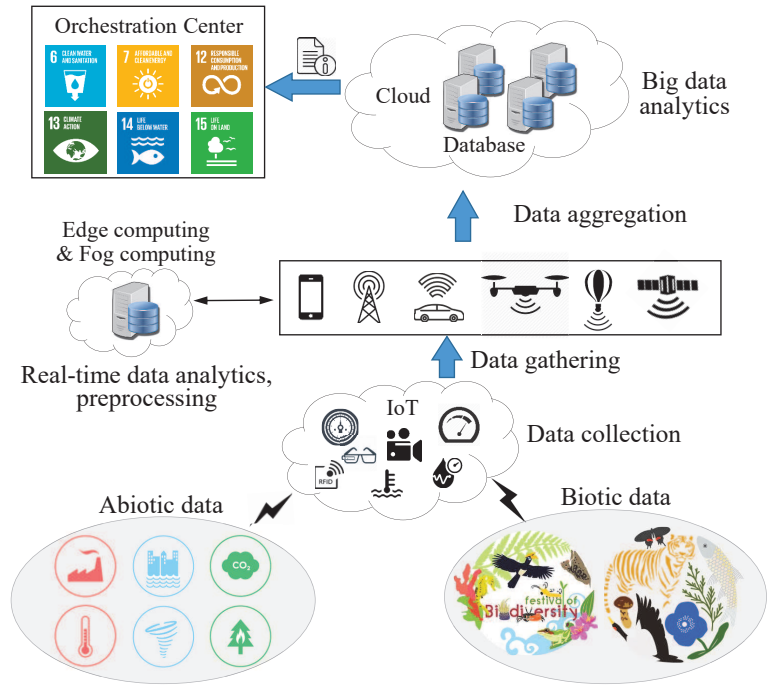

Fig. 5: The ICTs based architecture to promote the environmental SDGs.

[167] are in urgent need. Although the large volume of scientific data is available from various regional, national and global initiatives today, the efficient and effective utilization of these data to perform scientific analytics still remains as a big challenge. Therefore, the digitization and integration of biodiversity data are believed as the essential approaches to conserve environment and achieve the sustainable exploitation of natural resources. In addition, to efficiently manage and use these collected data, as well as to generate useful knowledge that can help make decisions and policies in the development of environmental sustainability, we need the support of the emerging powerful technological tools, such as service function chaining [168], edge computing [169], fog computing [170], and big data technologies [171], [172].

To this end, we present an ICT based orchestration architecture, which is shown in Fig. 5, for the promotion of environmental SDGs. Under this architecture, both the biotic and abiotic datasets can be sensed and collected by the IoT devices deployed pervasively. Then, the vehicular, aerial based mobile stations and mobile smart devices such as smartphones are used to gather the distributed data. The gathered datasets are either sent to the edge cloud for real-time data analytics and preprocessing, or sent to the remote cloud for big data analysis. Finally, the useful information extracted from the large volume of raw datasets is fed to the orchestration center for promotion of environmental SDGs. Meanwhile, to fulfill the appeal of the SDG 17 - partnerships for the goals, we are calling for research efforts relevant to the proposed architecture and case studies as well as real-world applications.

## VI. CONCLUSION

Although SDGs are directly relevant to so-called United Nations 2030 Development Agenda, SDGs actually would be still challenging issues in a long future human history, which should be far-far-far beyond 2030. Although UN only listed 17 SDGs in 2015, the number of global sustainability issues would be far-far-far more than 17. Those SDGs and targets would be strong guides toward the healthy development of global economy, societies, and environments. In this paper, we have focused on investigating the relevance between SDGs and ICT via panoramic reviews and discussions, and we have tried to provide our relevant understandings and visions on relevant issues. Although there have been a number of studies on several individual SDGs relevant to ICTS, overall speaking, in a number of aspects, there have been very limited research and development efforts for the investigations and solutions on how SDGs can be achieved or investigated based on or relevant to ICT, which would require many more global scientific, technological, industrial, and governmental efforts to promote and support the future achievements of SDGs. The impacts of SDGs on the activities of human beings would be positive, profound and significant. On one hand, ICT themselves have long term sustainability issues to be solved. On the other hand, ICT would have great potentials in playing important and key roles to support global economical, social, and environmental sustainability, which wait for the relevant discovery based on global and multi-disciplinary efforts.